# Denoising and Optical and SAR Image Classifications Based on Feature Extraction and Sparse Representation

BATTULA BALNARSAIAH, G RAJITHA

*Abstract-* Optical image data have been used by the Remote Sensing workforce to study land use and cover since such data is easily interpretable. Synthetic Aperture Radar (SAR) not only has the characteristic of obtaining images during all-day, all-weather but also provides object information that is different from visible and infrared sensors. However, SAR images have more speckle noise and fewer dimensions. This paper presents a method for denoising, feature extraction and compares classifications of Optical and SAR images. The image was denoised using K-Singular Value Decomposition (K-SVD) algorithm. An method to map the extraordinary goal signatures to be had within side the SAR or Optical image  using support vector machine (SVM) through offering given the enter facts to supervised classifier. Initially, the Gray Level Histogram (GLH) and Gray Level Co-occurrence Matrix (GLCM) are used for feature extraction.  Secondly, the extracted feature vectors from the first step were combined using the correlation analysis to reduce the dimensionality of the feature spaces. Thirdly, the Classification of SAR images were done in Sparse Representations Classification (SRC). The above-mentioned classifications techniques were developed and performance parameters are accuracy and Kappa Coefficient calculated using MATLAB 2018a.

## 1.1 Introduction

Remote Sensing has been recognized as a powerful device for deriving large-scale land cover information from satellite data. A great quantity of optical image data is likewise with no trouble to be had from earth observation satellites. Generation of images of land coverage started initially from aerial photographs, however for decades, spacecraft captured data have been used more often for determining characteristics of the information about land cover from high-resolution data such as LISS IV etc.. Land cover information is obtained for numerous scientific and engineering applications in defence, strategy arrangement or natural resources management purposes, etc. One of the main reasons for preference or wide usage of optical data is the relative ease of interpretation or understanding offered by the data.

Remote Sensing (RS) data are used majorly in Land Use Land Cover (LULC) classification application in which it is initiated to convert data into important information. LULC formations in two categories are water bodies and non-water bodies such as takes the fundamental role of change in earth and climate. In particular, in machine learning the supervised learning algorithms require large sets of training samples which is the vital difficulty of RS data. Spotlight on SAR and Optical technology, RS is capable of producing high resolution images using microwave by which it can capture water, settlements, and vegetation/agriculture and forest area.

A significant role is played by the synthetic Aperture Radar (SAR) Image classification in identifying the areas of the rural and urban, vegetation lands, buildings and also lakes etc. Necessary action is to be taken towards the automatic classification of SAR images. The SAR image is difficult to recognize the target in the midst of the beginning of the noise by the undesired situations as well as sudden disturbances [1]. Pixel-based, as well as Object-based classifications are the two techniques that are discovered within the image classification. Assigning every pixel to a single class in terms of feature space is the major objective of pixel-based image classification. Object-based classifications are grouping several pixels based on spatial properties of each pixel as they related to each other. Intensity and amplitude are the essential properties of image features [2]. The major problem in the SAR imagery is the occurrence of speckle noise, a signal-dependent granular noise, which degrades the quality, analysis and classification of images.

Hence, it is important to decrease the speckle noise from the SAR images before the image analysis starts, at the same time, the image details such as target points, textures, shapes should not be get disturbed. Recently, the speckle noise is reduced by proposing various conventional methods such as Lee[3], Frost[4], Kuan [5] as well as wavelet [6] filter etc. In recent times, image denoising has turned out to be a novel research hotspot that includes the Optical and SAR image despeckling based on K-SVD and classification with sparse representation.

## 1.2 Denoising Using K-SVD

Let x be the image patch of size $\sqrt{p} \times \sqrt{p}$ pixels, lexicographically ordered as a column vector of p length. It is assumed by the sparse representation model that x is built with the linear combination of s columns or also called as atoms, where $s \ll p$. This is deduced from the pre-specified over complete dictionary D, by applying DCT on the input raw image [7]. To give formally, $\mathbf{x} = \mathbf{D}\alpha$, where $\alpha \epsilon \mathbb{R}^m$ is a sparse vector with s non-zeros (this is denoted by $\|\alpha\|_0 = s$ ). Let us consider x, contaminated by noise v, with additive white speckle noise with zero mean and standard deviation $\sigma$. The MAP estimator can be obtained by solving

$$\hat{\alpha} = \arg\min_\alpha \|\alpha\|_0 \quad \text{s.t} \quad \|D\alpha - y\|_2^2 \leq p\alpha^2 \tag{3.1}$$

$\hat{x} = D\hat{\alpha}$, is the denoised image patch obtained with sparse representation vectors. The above expression optimized can be transformed to a lagrangian form below.

$$\hat{\alpha} = \arg\min_\alpha \lambda\|\alpha\|_0 + \frac{1}{2}\|D\alpha - y\|_2^2 \tag{3.2}$$

The above expression or equation is such that constraints become a consequence. With a proper choice of $\lambda$, it is signal (the vector y) dependent, the one or two problems can become equivalent. The local prior is imposed on every patch in X, where X is a large image of size $\sqrt{N} \times \sqrt{N}$ and Y is its noisy version. This gives the global MAP estimate after denoising the image [7].

$$\min_{\{\alpha_k\}_k, X} \frac{\mu}{2}\|\mathbf{X} - \mathbf{Y}\|_2^2 + \sum_k \left(\lambda_k \|\alpha_k\|_0 + \frac{1}{2}\|D\alpha_k - \mathbf{R}_k \mathbf{X}\|_2^2\right) \tag{3.3}$$

where, the first term the first term is the log-likelihood global force that demands a proximity between the measured image, **Y**, and its denoised (and unknown) version **X**. Put as a constraint, this penalty would have read $\|\mathbf{X} - \mathbf{Y}\|_2^2 \leq N\sigma^2$, which reflects the direct relationship between μ and σ.

The second term stands or the image prior that assures that in the constructed image, X, every patch, $x_k = R_k \mathbf{X}$ of size $\sqrt{p} \times \sqrt{p}$ in every location (thus, the summation by k) has a sparse representation with bounded error. The matrix $R_k \in \mathbb{R}^{p \times N}$ stands for an operator that extracts the $k^{th}$ block from the image. As to the coefficients $\lambda_k$, those must be spatially dependent, so as to comply with a set of constraints of the form $\|D\alpha_k - x_k\|_2^2 \leq p\sigma^2 k$ .

### 1.2.1 Obtaining the Dictionary D

The discussion so far has been based on the assumption that the dictionary D is known. This could be the case if we train it using the K-SVD algorithm over a corpus of clean image patches [8]. An interesting alternative is to embed the identification of D within the Bayesian formulation. Returning to the objective function in eq. (3.4) [9], the authors of also considered the case where D is an unknown,

$$\min_{\{\alpha_k\}_k, \mathbf{X}, \mathbf{D}} \quad \frac{\mu}{2} \|\mathbf{X} - \mathbf{Y}\|_2^2 + \sum_k \left( \lambda_k \|\alpha_k\|_0 + \frac{1}{2} \|\mathbf{D}\alpha_k - \mathbf{R}_k \mathbf{X}\|_2^2 \right) \quad (3.4)$$

In this case, **D** is learned using all the existing noisy patches taken from **Y** itself. Put more formally, a block-coordinate minimization is done: Initialize the dictionary **D** as the over complete DCT matrix and set **X** = **Y**. Then iterate between the OMP over all the patches and an update of the adaptive dictionary $D'$ using the K-SVD strategy [10]. The dictionary admits content, adapted to the image being treated, and the representations $\{\alpha_k\}_k$ are ready for a final stage in which the output image is computed via above equations.

## 1.3    Classification Methodologies

### 1. Support Vector Machine

In SAR or Optical data, example classifiers, classify the not unusual place instructions suitably, however extra intensive type is important for SAR or Optical mapping. Hence, it's far lots importance to have extra element spectral information. In order to split the elegance's expertise approximately spatial decision allocation is taken to locate the version within side the SAR or Optical of the identity elegance likewise. The training labels consequent are from 3 training which includes water bodies, settlements, Vegetation and Sand. In Neural Network flow toward, a given unknown pixel or a phase is classed into one of the pre-described training [11]. To keep away from unknown magnificence and to grind limitations among the support vector machine (SVM) is used. Three kernels of SVM are studied that allows you to locate excessive opportunity of correct results. Each supervised learning set of rules/ algorithm improves a reproduction of image that is used to categories/classifies novel labels within side the correct category. The SVM is specifically built to lessen a statistical sure at the simplification inaccuracy ensuing in models, it could extrapolate to new examples pretty well [12]. In addition to its probable, a randomized method is accompanied for fast mastering phase.

SVM is realistic through breaking the trouble down into some of binary problems. In this observe, 4 classes are in use wherein method of {4(4-1)}/2 classifiers are trained to discriminate among every pair of classes. Proceeding to class process, pixels of SAR or Optical images goals are received and used as training vectors which can be of high class quality statistics. This observes is studying the SVM kernels that are linear kernel. This kernels are represented as $(\vec{v} \cdot \vec{v}) = \vec{v} \cdot \vec{v}$ for linear. It is located that the linear SVM outperforms of the extra varieties of the SVM classifier. Figure three and 4 (b) offers the depicted evaluation of the linear the SVM classifier. The overall performance of SVM classifiers is evaluated the use of its preferred metrics, i.e., standard accuracy and K-coefficient. Measurement is primarily based totally on SAR or Optical images which imply the 4 categories/classes of RS (Remote Sensing) or image types.

### 1.3.1    Optical and SAR Image De-noising and Classification Using SRC

The block diagram shown below in figure 1 represents SAR image classification using Multi-Size patch-based SRC technique.

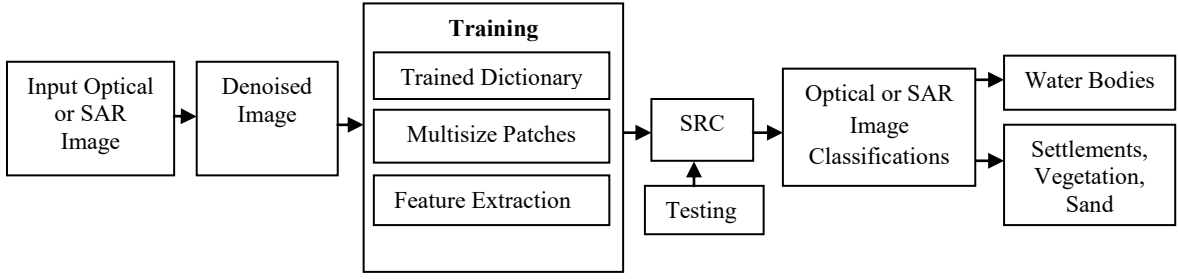

Figure:1 Block Diagram of Sparse Representation Classification (SRC) for SAR image

Denoising is applied on the input optical or SAR image first and multi-size patch-based dictionaries are created from the denoised image and further feature vectors are derived using GLH and GLCM, based on these feature vectors, a sparse representation classifier classifies the objects of the image and evaluates accuracy and Kappa Coefficient.

### 1.3.2 Sparse Representation Classification (SRC)

SRC represents linear subspace samples of a single class. Let if k classes are known, a dictionary can be defined by concatenating feature vectors of those classes. A series of training samples are used to formulate the testing sample y as shown below [13].

$$y = D\psi_0 \epsilon R^m \qquad (4)$$

Where, $\psi_0 = \left[0,,...0, \varphi_{i,1}, \varphi_{i,2}, \varphi_{i,n_i}, 0 ...,0\right]^T \epsilon R^n$ represents a coefficient vector, whose entries are non zero for the $i^{th}$ class, and the rest are zero entries. It is easier to estimate the identity of the testing sample y, when the sparsity of $\psi_0$ is more. The sparsest solution can be found by solving $y = D\psi_0$, based on this analysis the SRC problem is modeled as.

$$\hat{\psi}_0 = \text{argmin} \|\psi\|_0 \quad \text{subject to } D\psi = y \qquad (5)$$

Where, $\|\cdot\|_0$ represent the $l^0$ norm.

It is proven that, whenever solution $\psi_0$ is sparse enough, $l^0$-minimization problem equivalent to the $l^1$ minimization solution which is relaxed as

$$\hat{\psi}_1 = \text{argmin} \|\psi\|_1 \quad \text{subject to } D\psi = y \qquad (6)$$

Consider that $\delta_i: R^n \to R^n$ is the characteristic function on behalf of every single class.

The coefficient vector is represented by $\psi \epsilon R^n$ and the new coefficient nonzero vector is represented by $\delta_i(\psi)$. A linear reconstruction, $\hat{y}_i = D\delta_i(\hat{\psi})$ is approximated by the testing sample y [14]. The classification of y is done in accordance with the reduction of residual of $\hat{y}$ which is expressed below.

$$\min_i r_i(y) = \|y - D\delta_i(\hat{\psi})\|_2 \qquad (7)$$

The sparsity limited optimization problem in equation (4) is determined using OMP (orthogonal matching pursuit) which is a procedure for obtaining the sparse coefficients [15]. In order to recover the Sparse Signal support, the OMP Procedure is considered. A repetitive greedy method in which the column of the maximum correlated is selected using the current residuals and a combination of selected columns adds the selected column. The residuals are updated by the algorithms with the projection of observation within the linear subspace, that is stretched by the previously chosen columns and later it iterates the algorithm [16]. Simplicity in addition to fast implementation is the main advantages of OMP. In order to approximate and to recover a signal, this technique can be implemented. OMP algorithm is considered in this study that has a noise within it in the general settings. It is to be noted that residuals which are present in every single phase of the OMP algorithm possess orthogonality for the entire selected columns of X. Therefore, the selection of the column is never performed two times and it can be seen that at every single step, there is an increase in the set of selected columns. Stopping rule is an essential factor of an iterative procedure similar to OMP. The bounded noise as well as Gaussian noise cases provides certain stopping rules on behalf of the OMP algorithm.

### 1.3.3 Multi-Size Patches Based on Dictionary

Construction of a dictionary is the initial step to be done in sparse representation application, where its construction cannot be done directly using the pixel values [17]. The transformation of the pixel value space within a feature space is essential for overcoming the topography as well as geomorphology's complexity. Due to which, the discriminative features are extracted out of the optical or SAR image with less computational cost. The feature vectors are derived by exploiting the statistical characteristics within the gray-level histogram (GLH) as well as the texture statistics within the GLCM. In order to represent every single pixel, a concluding feature vector is formed by concatenating two vector types [18]. With the representation of the statistical data and by capturing texture data within adjoining regions, a competitive performance is provided by the nonlinear feature. Assume $n_i$ vectors of training samples are selected from the $i^{th}$ class within the $h^{th}$ layer as columns for constructing a matrix $A_i^h = [x_{i,1}^h, x_{i,2}^h, ..., x_{i,n_i}^h] \epsilon R^{m \times n_i}$ where, the feature vector dimension is denoted by m. The definition of dictionary $D^h$ is given by means of the concatenating the $n_i$ training sample vectors concerning the entire K classes that are specified within the fixed $h^{th}$ layer [13].

$$D^h = [A_1^h, A_2^h, ..., A_K^h] = [x_{1,1}^h, x_{1,2}^h, ..., x_{K,n_K}^h] \quad (8)$$

Where, a fixed hierarchical dictionary is denoted by $D^h$. Let $h = \{h_1, h_2, ... h_H\}$ accounts an H latent layers; moreover it also denotes the overall classification number. The pixel points are represented by a fixed h-level patch by extracting the feature vectors and a fixed-level dictionary $D_h$. The definition of patch size (odd) can be represented as $S_h = \{S_1, S_2, ..., S_N\}, S_1 > S_2 > \cdots > S_N$. Where $S_h$ represents size and N represents the number of patches. The patch sizes and the number of layers are related to each other as stated previously and however they are not individually correspondent one to one all the time. The working of human visual system and the multi-size patch-based method is similar, where precise observation is made once after the initial glance at the outline of the object. Here, the feature of every single pixel is extracted by using three dimensions. Consider that the resolution of the image is represented by PS. Initially, a size $S_{init} = 2^* floor\left(\frac{PS*3}{2}\right) + 1$ is selected and later a suitable largest odd patch size $S_{large}$ is selected, depending upon various types of terrain and the size of the text on in the neighborhood of $S_{init}$. Next, $3 \times 3$ is set as the smallest number of patch size $S_{small}$. And the centre patch size is chosen as given by the equation below

$$S_{middle} = S_{small} + (S_{large} - 3)/2 + mod\left((S_{large} - 3)/2, 2\right) \quad (9)$$

### 1.3.4 Multiple SRC

In optical or SAR image classification, since SAR image contains various complex terrains; it is difficult to have effective training samples to represent each pixel. To overcome this and to improve the optical or SAR image classification, sparse representation classifier with double threshold is used. The issue of sparse representation problem is described with $\ell^0$ minimization in equation 3, which is solved approximately by greedy pursuit algorithm (GPA) [19].

In this, OMP algorithm uses the support atoms and calculates the representation coefficients. Further reliability of the classified pixels is estimated based on the judgment on the residual $r_i(y)$. If it is within a specified range, we treat it as reliable pixel, and otherwise it is assumed as uncertain one. This may be calculated in the after that layer. The training of the samples are extracted from the labelled points of the previous classification results in the next classification process. Calculations of the support atoms and for calculating the representation coefficients, this method implement OMP algorithm. The accuracy of classified pixels is estimated by judging upon the residual $r_i(y)$. Specifically, the definition of pixel is done by means of a reliable one whenever $r_i(y)$ lies within a particular range. The definition of the pixel is done alternatively by means of an unreliable one and its determination is done in the other level. The labeled points within the outcomes of the earlier classification are used to extract the training samples within the next classification. The selection of the training samples within the SAR image is done in a random way. The traditional SRC is used to classify the uncertain data points in the final layer [20]. The expression for minimum residual $r_{min}^h(y)$ is given below.

$$\tau^h = r_{min}^h(y) = \min_{i=1...K} r_i^h(y)$$

$$\theta^h = arg \min_{i=1...K} r_i^h(y) \quad (10)$$

Where, the sum of the class & label using the least residual are denoted by K and $\theta$. The minimum $r_i(y)$ is represented as $\tau^h$ for eliminating the ambiguity in the algorithm given below. The determination of labeling every single pixel below the residual judgment is expressed as:

$$\text{Label}(y)^h = \begin{cases} \min_i r_i^h(y) \\ \text{uncertain} \end{cases}, \tau^h \leq \Delta_1, \left|r_{i \neq \theta^h}^h(y) - \tau^h\right| \geq \Delta_2 \quad (11)$$

Where, the thresholds are represented as $\Delta_1$ and $\Delta_2$. Two kinds of effective messages are conveyed by two terms of judgments. The smaller the reconstruction error $r_{min}^h(y)$ is represented by the first term $\tau^h \leq \Delta_1$ is more similar to y and class $\theta$. Under a smaller residual, class $\theta$ is used to represent y effectively sparsely. The second term represents $\left|r_{i \neq \theta^h}^h(y) - \tau^h\right| \geq \Delta_2$ and y can be efficiently represented by class $\theta$. Moreover, a strong constraint is added by the two terms for the interclass distance and intra-class distance which means that this situation is related to class $\theta$. The uncertain pixels decrease step-by-step by increasing h. Here, until the entire dictionaries are used, the usage of the multisize patch-based dictionary is done often.

Algorithm .1 Multi-size patches on the basis of numerous SRC.

1. Construction of initial dictionary (h=1): considering K classes, choose $n_i$ samples out of every single class in a random way, every sample is denoted using m-dimensional vector which compresses the initial dictionary $D^{init} = [A_1^{init}, A_2^{init}, \ldots, A_K^{init}]$, init = 1, in which the extraction of the feature vectors is done with the first layer patches.
2. Multiple classifications ($1 \leq h \leq H - 1$: categorize the entire pixels with the help of equation 8 in the initial layer in addition to undefined opinions within additional layers. Identify the classification of testing sample y.
3. Construction of $h^{th}$ dictionary ($2 \leq h \leq H$): Select n points are selected out of every single category that is characterized as novel the training samples within the earlier layer, and a feature vector $x_{i,j}^h$ at h-th layer is extracted. The hierarchical dictionary $D^h = [A_1^h, A_2^h, \ldots, A_K^h] \in R^{m \times n}$ is constructed by arranging the vectors as columns.

Last layer Classification (h = H): with the help of traditional sparse representation classifier, the uncertain points in SAR image are classified using the $H^{th}$ dictionary $D^H$.

**Training of Feature Extraction Based on Multisize Patches:**

As a result of the imaging mechanisms that differ from the nature imagery, it is not possible to apply SRC to the classification of SAR images. The application of the SRC within the SAR images can be done in an effective way, whenever the transformation of the SAR image is done within a specific characteristic space [21]. In this chapter the SAR image characteristics by implementing a histogram of gray-level, in addition to a gray-level co-occurrence matrix (GLCM) [22]. Objects are distinguished and characterized by the stability that is possessed by the Gray-level histogram against the noise in addition to the variation in the perception. The various characteristics of GLCM are on the basis of the second order statistics [23]. With respect to the various concepts of homogeneity, dissimilarity, and uniformity contrast and so on, the complete average for the correlation degree among the pair of pixel are reflected by using it. The difference between the pixels is a primary component, which affects the GLCM's discriminating abilities. The degree of correlation among the neighbouring pixels (short-distance neighbourhood connectivity) is reflected by considering the initial distance and by increasing the value of the distance. Thus, an efficient distribution component is extracted by the SAR image and moreover, it is implemented within various functions of the classification. A range of various size patches are utilized for capturing the micro textures as well as macro textures in an appropriate manner, and represented using GLCM within an original SAR image. Therefore, the gray-level histogram in addition to GLCM are joined together to form a discriminative feature vector.

In our proposed algorithm, the SAR pixels can be sparsely represented as a linear adaptive or combination of a dictionary $D'$, which is consist of a number of training samples chosen from the training set, then we can classify SAR via sparse representations of the original pixels. Unlike most existing SR algorithms via sparse representation, in this SRC an algorithm is proposed. The main contributions lie in three aspects, 1) Denosing of the SAR image using K-SVD 2) The learning the dictionary is DCT with K-SVD algorithm is used to obtain a more adaptive dictionary $D'$ 3) When solving the sparse representation of testing pixels, the contextual information is joined into the sparse model to obtain more optimized sparse representation of pixels to improve the classification performance [24]. The experimental results show that our approach yields good classification results on the well-known public SAR image of the selected data or SAR image with only two simple linear classifiers are water bodies and Non-water bodies like settlements, vegetation and open land.

In this work 256 Trained DCT Dictionary of image and Multisize patches, appropriate matches occurred for optical is 212 labels/patches corrected and second one is SAR image 180 matches/label appropriate corrected. The DCT and Multisize matches are as shown in figure 2 below.

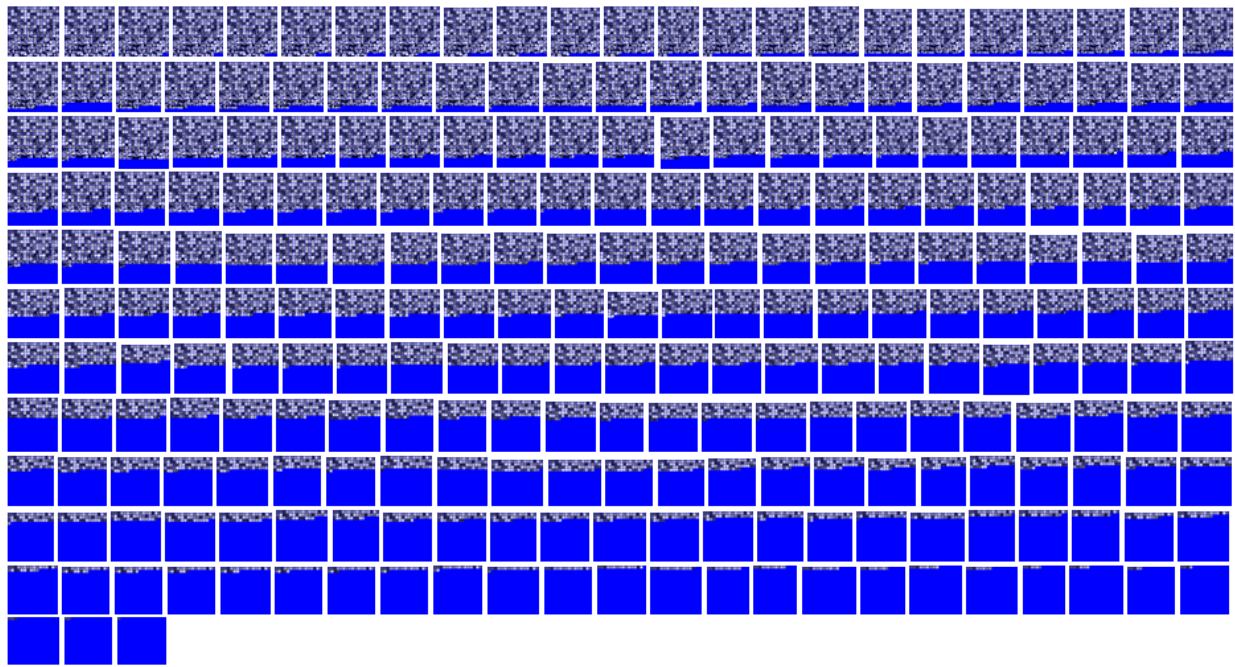

Figure 2: DCT Trained and Multisize Patches

## 1.4 Results and Discussion

The illustrated reference shows in figure 3 (a) Original optical image of size 279x375, (b) DE-noised Optical image, (c) trained DCT dictionary and (d) Multisize patches of K-SVD and its adaptive Dictionary $D'$, (e) Classified Optical Image using SVM (f) Classified Optical image using SRC. Figure 4 (a) Original SAR image of size 411x453, (b) De-noised SAR image, (c) trained DCT dictionary and (d) Multisize Patches of K-SVD and its adaptive Dictionary $D'$ (e) Classified SAR Image using SVM (f) Classified SAR image using SRC. The illustrated results of the Optical and SAR de-noised and classified of the images comparison in figure 5.

The optical and SAR image 279x315, 411x453 sizes are classified using SVM by qualified information of the numerous based on SVM linear kernel methods by referring both the images and an addition to the classification of the optical and SAR image of size 279x315 and 411x453 using multisize patches based SRC methodology.

The overall optical and SAR images of noised and de-noised comparison parameter as shown in Table 1

Table.1 Comparison of noise and de-noise of the Optical and SAR Images

| S.N | Name of the Parameter | Optical Image | SAR Image |
|---|---|---|---|
| 1 | Image Size | 279x315 | 411x453 |
| 2 | Number of Atoms in the Dictionary | 256 | 256 |
| 3 | Maximum coefficient for each Signal | 3 | 3 |
| 4 | Noise Level(Sigma) | 10 | 20 |
| 5 | PSNR Noise in dB | 34.08 | 37.10 |
| 6 | PSNR Denoise in dB | 5.99 | 13.73 |
| 7 | Number of Patched/Label Corrected | 232 | 237 |

$$Accuracy = \frac{Matched\ Labels\ (Patches) overall\ the\ class}{Total\ number of Labels\ (Patches) overall\ the\ class} \times 100 \qquad (12)$$

**The K-Coefficient**

The K-coefficient is a measure of integrator agreement which is given as

$$K = \frac{P_{ps} - P_{ups}}{1 - P_{upse}} \qquad (13)$$

Here, $P_{ps}$ predicted pixel samples and $P_{upse}$ unpredicted pixel sample, Kappa (K) is a positive value by means of its magnitude reflecting the strong point of the integrator conformity and it becomes negative when the experiential conformity is less than the chance conformity. To bring the best and appropriate kernel for SVM, dissimilar kernels have been considered to find the reliable one and also other method to utilize.

An exported the both image classifications are evaluated using SVM of accuracy and kappa coefficient 86.74%, 88.49% and 0.8568, 0.8795 respectively. It is worth noting that the entire classification of Optical and SAR image 90.62% and 92.57% accuracies and 0.8974, 0.9186 kappa coefficient of optical and SAR respectively are in Table 2 attained using the various patches or labels across the representation of the sparse. The accuracies are evaluated as in above equation is 12

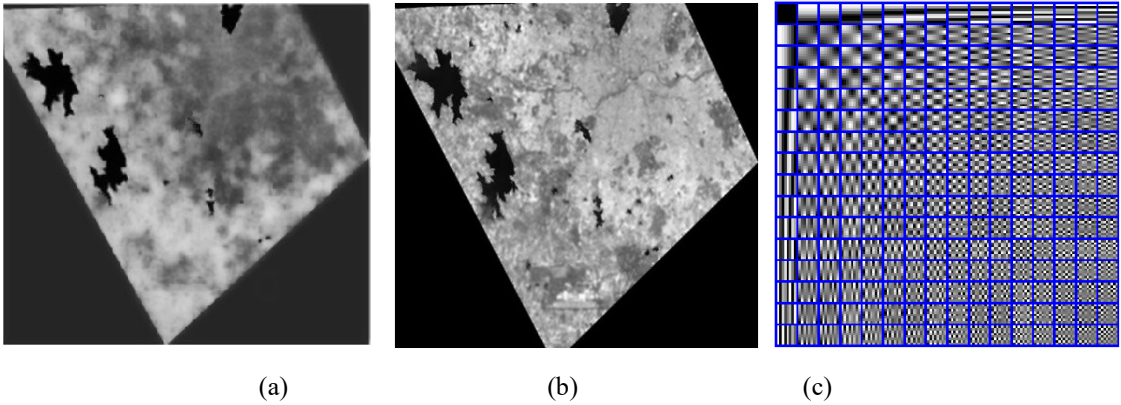

(a)            (b)            (c)

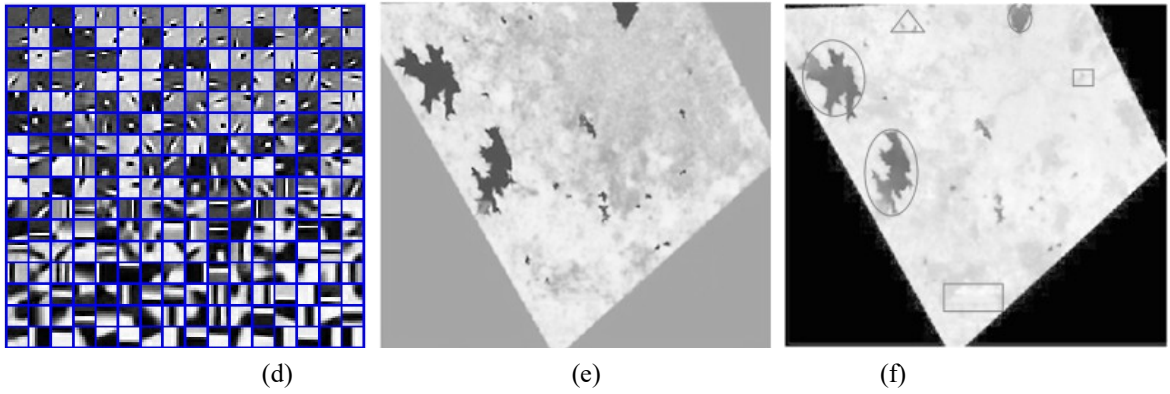

Figure 3 (a) Original Optical Image size 279x315 (b) De-Noisy Image (c) Trained DCT Dictionary of image (d) Multi-Size patches (e) Classified Optical Image using SVM (f) Classified Optical Image using SRC

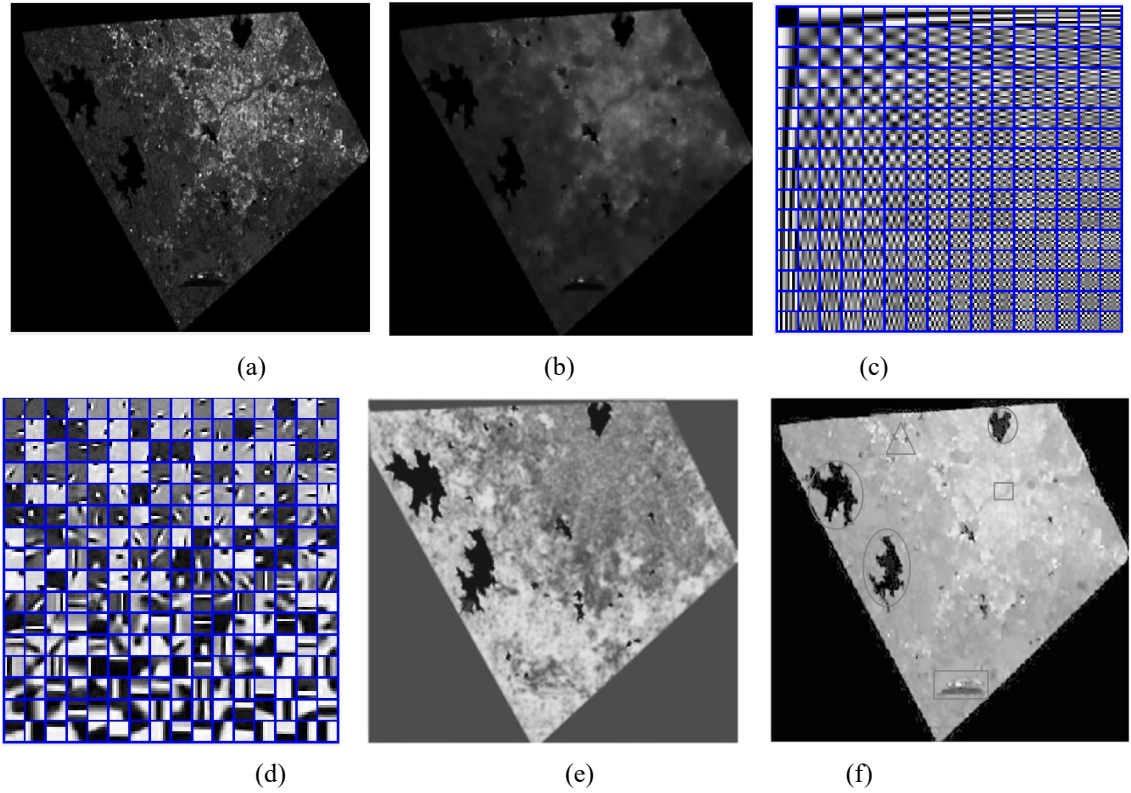

Figure 4 (a) Original SAR Image size 411x453 (b) Noisy Image (c) Trained DCT Dictionary of image (d) Multisize patches (e) Classified SAR Image using SVM (f) Classified SAR Image using SRC

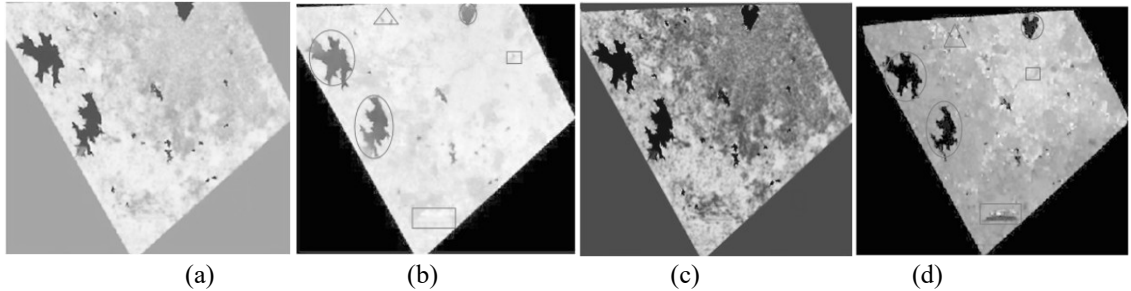

(a)          (b)          (c)          (d)

Figure: 5 Comparison of the Optical and SAR classified images using SVM (6. (a) And (c), SRC using (b) and (d)

◯ : Water Bodies      ▭ : Settlements      ▫ : Sand      △ : Vegetation

Table: 2 Classification of Optical and SAR Image Accuracies

| S.No | Classifications Techniques | Type of Images | Labels/Patches | Accuracy% | Kappa Coefficient |
|---|---|---|---|---|---|
| 1 | SVM | Optical | Based on Kernels classified | 86.74 | 0.8568 |
|   |     | SAR | Based on Kernels classified | 88.49 | 0.8795 |
| 2 | SRC | Optical | Total Number of Labels(Patches) over all the Class=256<br>Matched Labels(Patches) over all the Class=232 | 90.62 | 0.8974 |
|   |     | SAR | Total Number of Labels(Patches) over all the Class=256<br>Matched Labels(Patches) over all the Class=237 | 92.57 | 0.9186 |

## 1.5 Conclusion

In this paper to assess denoising procedures, an after de-noised SAR and Optical image, first methodology is using support vector machine classified the SAR and Optical image classifications. The second methodology is SRC for both SAR and Optical image classifications. The SAR and Optical the classification based on the kernel techniques of SVM. The training of the multiple patches is done across both the Optical and SAR image classification using SRC technique. Subsequently, using trained information of several patches, the Remote Sensing (RS) data or image is four classifications are water bodies, settlements and another classes is vegetation and sand using both the techniques or algorithms that is SVM and SRC. Both the Optical and SAR classifications are using SRC, attained using the various patches across the representation of the sparse representation. The SRC is the most significance than the SVM in remote sensing of Optical and SAR image classifications. The work carried out indicates the SRC efficiency in image classifications.

**Acknowledgment**

The authors would like to thank all reviewers, General Manager of Outreach Facility of NRSC (ISRO), Hyderabad for their encouragement. The authors would also like to extend their sincere thanks to the institute staff for the technical support and remarkable suggestions during research work. The author would like to acknowledge CSIR, the fellowship provided by Govt. of India. New Delhi, India.